\documentclass[copyright,creativecommons]{eptcs}
\providecommand{\event}{ACL2 2025} 
\usepackage{breakurl}             
\usepackage{underscore}           

\usepackage{color}

\usepackage{fancyvrb}

\usepackage{float}
\floatstyle{boxed}
\restylefloat{figure}

\newcommand{\hrefdoc}[2]{\href{https://www.cs.utexas.edu/users/moore/acl2/manuals/latest/index.html?topic=#1}{\underline{#2}}}

\newcommand{\hrefdoctt}[2]{\hrefdoc{#1}{\texttt{#2}}}

\definecolor{darkcyan}{cmyk}{1.0,0.2,0.2,0.2}

\title{Extended Abstract:  Partial-encapsulate and Its Support for Floating-point Operations in ACL2}

\author{Matt Kaufmann and J Strother Moore
\institute{Department of Computer Science,
The University of Texas at Austin, Austin, TX, USA
(retired)}
\email{\{kaufmann,moore\}@cs.utexas.edu}
}
\def\authorrunning{M.~Kaufmann \& J S.~Moore}

\def\titlerunning{FP in ACL2}

\begin{document}
\maketitle




\section{Introduction}

The
\hrefdoctt{ACL2\_\_\_\_PARTIAL-ENCAPSULATE}{partial-encapsulate}\footnote{Underlined
  links are to ACL2 documentation topics.} macro was introduced in
ACL2 Version 8.2 (May, 2019), providing a general way to evaluate
constrained functions, thus generalizing
\hrefdoc{ACL2\_\_\_\_DEFINE-TRUSTED-CLAUSE-PROCESSOR}{trusted
  (unverified) clause-processors}~\cite{clauseprocessors}.  However,
the ACL2~\hrefdoc{ACL2\_\_\_\_COMMUNITY-BOOKS}{community
  books}~\cite{community-books} of ACL2 Version 8.6 contain only a few
applications of this utility.  One goal of this extended abstract is
to publicize (finally) this powerful utility.  We do so by describing how
it supports floating-point (FP) computation in ACL2, which addresses
our second goal: to augment the very brief discussion of that support
in our published treatment of FP computation in
ACL2~\cite{fp-jones-festschrift}.

This extended abstract is intended to be reasonably self-contained,
especially when combined with the supporting materials described
below.  For much more background on FP computation in ACL2, see
\hrefdoc{ACL2\_\_\_\_DF}{its documentation topic} for user-level
discussion; and for implementation-level comments, see the ACL2 source
code, especially file \texttt{float-a.lisp} and the comment therein,
{\em Essay on Support for Floating-point (double-float, df) Operations
  in ACL2}.

FP computations are widely used in the scientific community, and they
are generally much faster than computations with rationals.  ACL2
supports such computations using {\em double-floats} (FPs), which are
a Lisp\footnote{In this paper, ``Lisp'' refers to Common
  Lisp~\cite{cl-hyperspec}.} datatype typically consisting of
double-precision floating-point numbers.  But FP operations are
awkward to axiomatize.  The following Lisp computations show that FP
addition is not associative (which is awkward since ACL2 \texttt{+} is
axiomatized to be associative) and in Lisp, the \texttt{EQUAL}
function does not compute equality on numbers.

\begin{Verbatim}[commandchars=\\\{\}]
? (setq *read-default-float-format* 'double-float){\em ; read FPs as double-floats}
DOUBLE-FLOAT
? (+ 0.1 (+ 0.2 0.3))
0.6
? (+ (+ 0.1 0.2) 0.3){\em ; not the same result as above; associativity fails!}
0.6000000000000001
? (equal 1 1.0){\em ; two equal arithmetic values need not satisfy EQUAL}
NIL
? 
\end{Verbatim}

A solution might be to add a new FP datatype to the ACL2 logic, but we
were loath to complicate ACL2 that way.  In particular, although that
could explain a result of \texttt{nil} for the evaluation of
\texttt{(equal 1 1.0)}, it would be at odds with a result of
\texttt{t} for the evaluation of \texttt{(= 1 1.0)}, since \texttt{=}
is defined logically to be \texttt{EQUAL}.  The new datatype would
also probably complicate ACL2's
\hrefdoc{ACL2\_\_\_\_TYPE-REASONING}{type reasoning}.

Instead, ACL2 models FPs as the rational numbers they {\em represent};
a rational is {\em representable} if it is the numeric value of a
double-float.  By tracking the use of FP expressions much as
\hrefdoc{ACL2\_\_\_\_STOBJ}{stobjs}~\cite{stobjs} are tracked, ACL2
arranges for Lisp FP computations to be performed using Lisp
double-floats, even though they are rational operations logically.

Our goal is to illustrate how \texttt{partial-encapsulate}, when combined with
redefinition in Lisp allowed by a \hrefdoc{ACL2\_\_\_\_DEFTTAG}{trust
  tag}~\cite{ttag}, can extend the power of ACL2.
We illustrate this idea by showing a toy example that supports FP operations.  For simplicity, this
exposition ignores the stobj-like tracking mentioned above; as a
result (and as noted at the end below), this toy implementation is
actually unsound!  That observation highlights the potential danger of
using \texttt{partial-encapsulate} together with redefinition in Lisp.
The actual ACL2 implementation of floating-point operations also uses
\texttt{partial-encapsulate} but avoids unsoundness by taking great
care, including the use of stobj-like tracking mentioned above.

Although our toy example involves floating-point numbers, we expect
most user applications would avoid data types not supported by the ACL2
logic (like floating-point).  That should make it considerably less
complicated to avoid unsoundness than was the case when adding support
to ACL2 for FP operations.

\section{A Toy Implementation Illustrating {FP} Support}

We describe the example worked out in the supporting materials for
this paper, which can be found in the following files in community
books directory \texttt{books/demos/fp/}.

\begin{itemize}

\item \texttt{fp.lisp} --- Certifiable book introducing some FP operations logically

\item \texttt{fp-raw.lsp} --- Lisp redefinitions supporting FP computation

\item \texttt{fp.acl2} --- Certification support for trust tag and dependencies

\end{itemize}

\noindent They define a few functions with ``fp'' in the name, which
correspond to analogous ACL2 built-ins with ``df'' (for
``double-float'') in the name instead of ``fp''.  (There are many more
\hrefdoc{ACL2\_\_\_\_DF}{df} built-ins as well.)  Square root and
addition functions are introduced logically in \texttt{fp.lisp} using
\texttt{partial-encapsulate} but are given executable Lisp definitions
in file \texttt{fp-raw.lsp}.  To support these, we also introduce a
conversion function \texttt{to-fp} and a recognizer function
\texttt{fpp} in \texttt{fp.lisp}, as follows.  Think of \texttt{(to-fp
  x)} as choosing a representable rational near \texttt{x};
specifically, it chooses the rational returned by evaluating the
expression \texttt{(float x 0.0D0)} in Common Lisp, as discussed
further below.

\begin{Verbatim}[commandchars=\\\{\}]
(partial-encapsulate{\em ; introduce conversion to representable rationals}
 (((constrained-to-fp *) => * :formals (x) :guard (rationalp x)))
 nil{\em ; supporters; see documentation for partial-encapsulate}
 (local (defun constrained-to-fp (x) (declare (ignore x)) 0))
 (defthm rationalp-constrained-to-fp
   (rationalp (constrained-to-fp x))
   :rule-classes :type-prescription)
 (defthm constrained-to-fp-idempotent
   (equal (constrained-to-fp (constrained-to-fp x))
          (constrained-to-fp x)))
 ... {\em ; other exported defthm events omitted here}
)
(defun to-fp (x){\em ; convert to representable rationals}
  (declare (xargs :guard (rationalp x)))
  (constrained-to-fp x))
(defun fpp (x){\em ; recognizer for representable rationals}
  (declare (xargs :guard t))
  (and (rationalp x) (= (to-fp x) x)))
\end{Verbatim}

A \texttt{partial-encapsulate} event represents a corresponding,
implicit \texttt{encapsulate} event that introduces additional
exported theorems.  The key requirement is that the axioms exported by
that event, including the implicit additional ones, are all provable
for some choice of local witnesses for the signature functions.  See
the documentation topic for
\hrefdoc{ACL2\_\_\_\_PARTIAL-ENCAPSULATE}{partial-encapsulate} for
more information about that utility, in particular its lack of support
for \hrefdoc{ACL2\_\_\_\_FUNCTIONAL-INSTANTIATION}{functional
  instantiation} due to unknown constraints.

In the case of \texttt{constrained-to-fp}, the implicit constraints
(from additional, hidden \texttt{defthm} events) include a theorem for
each computation result based on the following definition from
\texttt{fp-raw.lisp}; for example, since \texttt{(float 1/3 0.0D0)}
computes to an FP with value \\ 6004799503160661/18014398509481984, an
implicit axiom is \\ \texttt{(equal (to-fp 1/3)
  6004799503160661/18014398509481984)}.

\begin{verbatim}
(defun to-fp (x)
  (declare (type rational x))
  (float x 0.0D0))
\end{verbatim}

\noindent Of course, there are in principle infinitely many such
implicit axioms.  But the implicit \texttt{encapsulate} event is a
finite object, so we consider only computation results that will be
performed, somewhere by someone, using the current version of ACL2.
For details, see comments in the \texttt{partial-encapsulate} that
introduces function symbol \texttt{constrained-to-df} in ACL2 source
file \texttt{float-a.lisp}.

Why don't we instead introduce \texttt{to-fp} with
\texttt{partial-encapsulate} and eliminate the function
\texttt{constrained-to-fp}?  The reason is that the ACL2 rewriter
refuses to execute calls of constrained functions (regardless of
redefinition in Lisp).  This way, ACL2 succeeds, for example, in the
proof of \texttt{(thm (equal (to-fp 1/4) 1/4))}.

The function \texttt{fp-round} is similar to \texttt{to-fp}, but these
two functions serve different purposes.  \texttt{To-fp} is intended to
be executable.  \texttt{Fp-round}, which is not executable, logically
supports defining FP addition to be the rounded result of exact
addition, as specified by IEEE Standard 754~\cite{IEEE-754-2019}.  FP
addition is defined as follows in \texttt{fp.lisp}.

\begin{verbatim}
(defun fp+ (x y)
  (declare (xargs :guard (and (fpp x) (fpp y))))
  (fp-round (+ x y)))
\end{verbatim}

\noindent \texttt{Fp+} is redefined in \texttt{fp-raw.lsp} as follows.
Note that for FPs x and y, the Lisp \texttt{+} operation does the
requisite rounding.

\begin{verbatim}
(defun fp+ (x y)
  (declare (type double-float x y))
  (+ x y))
\end{verbatim}

For more details see the aforementioned supporting materials, which in
particular contain:

\begin{itemize}

\item redefinition in Lisp using a trust tag followed by the form
  \texttt{(\hrefdoc{ACL2\_\_\_\_INCLUDE-RAW}{include-raw}
    "fp-raw.lsp")} in \texttt{fp.lisp}, to load \texttt{fp-raw.lsp}
  into Lisp, which redefines functions already defined in ACL2;

\item introduction of the FP square root function,
  \texttt{fp-sqrt}, using \texttt{partial-encapsulate} for ACL2 and
  Lisp \texttt{sqrt} for execution;

\item handling of
  \hrefdoc{ACL2\_\_\_\_EVALUATION}{executable-counterpart} (so-called
  ``*1*'') functions for redefined functions;

\item tests showing that evaluation works, even during proofs; and

\item examples demonstrating the need for care when using Lisp
  redefinition, by proving \texttt{nil}.

\end{itemize}

\noindent The soundness issue just above is due to the attempt to
traffic in a Lisp datatype (double-float) that is not supported in the
ACL2 logic.  Comments in \texttt{fp.lisp} outline how ACL2 avoids
these problems for its \hrefdoc{ACL2\_\_\_\_DF}{df} implementation.
We expect that most user applications of \texttt{partial-encapsulate}
can avoid such soundness issues if appropriate care is taken.

\paragraph{Acknowledgments.}  We thank Warren Hunt for encouraging
the implementation of floating-point operations in ACL2 and
ForrestHunt, Inc. for supporting that implementation.  We also thank
the reviewers for helpful comments.

\bibliographystyle{eptcs}
\bibliography{fp}
\end{document}